\documentclass[manuscript]{aastex}

\def\pmb#1{\setbox0=\hbox{#1}
  \kern-.02em\copy0\kern-\wd0
  \kern.01em\copy0\kern-\wd0
  \kern.01em\copy0\kern-\wd0
  \kern.01em\copy0\kern-\wd0
  \kern.01em\copy0\kern-\wd0
  \kern-.02em\raise.01em\box0 }

\def\ref#1#2{$^{#1}$}

\shorttitle{Age of HD 140283}
\shortauthors{Bond et al.}

\begin{document}

\title{Episodic Mass Loss from the Hydrogen-Deficient Central Star of the
Planetary Nebula Longmore~4\altaffilmark{1}}

\author{Howard E. Bond\altaffilmark{2,3}
}

\altaffiltext{1}
{Based on observations with the  1.5-m telescope operated by the SMARTS
Consortium at Cerro Tololo Interamerican Observatory.}

\altaffiltext{2}
{Space Telescope Science Institute, 
3700 San Martin Dr.,
Baltimore, MD 21218, USA}

\altaffiltext{3}
{Current address: Department of Astronomy \& Astrophysics, Pennsylvania State
University, University Park, PA 16802, USA; heb11@psu.edu}

\begin{abstract}

A spectacular transient mass-loss episode from the extremely hot,
hydrogen-deficient central star of the planetary nebula (PN) Longmore~4 was
discovered in 1992 by Werner et al.  During that event, the star temporarily
changed from its normal PG~1159 spectrum to that of an emission-line
low-luminosity early-type Wolf-Rayet [WCE] star. After a few days, Lo~4 reverted
to its normal, predominantly absorption-line PG~1159 type. To determine whether
such events recur, and if so how often, I monitored the optical spectrum of Lo~4
from early 2003 to early 2012.  Out of 81 spectra taken at random dates, four of
them revealed mass-loss outbursts similar to that seen in 1992. This indicates
that the episodes recur approximately every 100~days (if the recurrence rate has
been approximately constant and the duration of a typical episode is
$\sim$5~days), and that the star is in a high-mass-loss state about 5\% of the
time. Since the enhanced stellar wind is hydrogen-deficient, it arises from the
photosphere and is unlikely to be related to phenomena such as a binary or
planetary companion or infalling dust. I speculate on plausible mechanisms for
these unique outbursts, including the possibility that they are related to the
non-radial GW Vir-type pulsations exhibited by Lo~4.  The central star of the PN
NGC~246 has stellar parameters similar to those of Lo~4, and it is also a
GW~Vir-type pulsator with similar pulsation periods. I obtained 167 spectra of
NGC~246 between 2003 and 2011, but no mass ejections were found. 


\end{abstract}



\section{Introduction}

Longmore~4 (Lo~4; PN~G274.3+09.1) is a low-surface-brightness planetary nebula
(PN), discovered by Longmore (1977) during the ESO-SRC southern-hemisphere sky
survey. Spectroscopic observations of its $V=16.6$ central star by M\'endez
et~al.\ (1985) led to a spectral classification as a PG~1159 star.

PG~1159-type objects (the name refers to the prototypical star PG~1159$-$035)
are extremely hot, hydrogen-deficient planetary-nebula nuclei (PNNi) and white
dwarfs, having spectra with a conspicuous \ion{C}{4} and \ion{He}{2} absorption
trough at 4659--4686~\AA, and no features due to hydrogen (apart from a rare
class of ``hybrid PG~1159'' stars showing traces of H)\null. The high-excitation
\ion{O}{6} 3811-3834~\AA\ doublet is often seen in emission in the spectra of
these stars, and if so they are sometimes referred to as having
``\ion{O}{6}''-type spectra, or early~WC types.  Werner, Rauch, \& Kruk (2007)
found atmospheric parameters for the Lo~4 central star of $T_{\rm
eff}=170,000$~K, $\log g=6$. This result, based on the detection of \ion{Ne}{8}
absorption lines in {\it FUSE\/} FUV spectra, was a substantial upward revision
in the temperature compared to an earlier analysis that had given $T_{\rm
eff}=120,000$~K, $\log g=5.5$ (Rauch \& Werner 1997).

The spectrum of Lo~4 is quite similar to that of K~1-16, the first known
pulsating PNN (Grauer \& Bond 1984), and a member of the GW~Vir class of
non-radially pulsating white dwarfs (reviewed in recent years by Fontaine \&
Brassard 2008, Quirion 2009, and Althaus et al.\ 2010). This close similarity
was borne out by the discovery of photometric pulsations in Lo~4 by Bond \&
Meakes (1990). Lo~4 was, at the time, only the second known pulsating PNN,
having a strong periodicity near 31~min (1850~s), along with at least 8 other
pulsation modes ranging from 831 to 2325~s. Based on its light variations, the
central star has been given the variable-star designation LV~Velorum, but in
this paper I will refer to the star as Lo~4.

\section{A Spectacular Mass-Loss Event in Lo 4}

A remarkable transient mass-loss event that occurred in Lo~4 in 1992 was
discovered serendipitously by Werner et~al.\ (1992, 1993). For several days the
spectrum changed from PG~1159 to a low-luminosity early carbon Wolf-Rayet, or
[WCE], type of about [WC2-3]. During this event, the spectrum of Lo~4 had strong
emission at the \ion{C}{4} complex near 4659~\AA, at \ion{He}{2} 4686~\AA, and
at \ion{O}{6} 5291~\AA\null. Within a few days, the star reverted back to its
usual PG~1159 spectrum, with the \ion{C}{4} and \ion{He}{2} features once more
in absorption. The mass-loss rate during this outburst was estimated at $\dot M
\simeq 5\times10^{-8} \, M_\odot \, \rm yr^{-1}$  by Werner et al.\ (1993). As
emphasized by Werner et al., this dramatic and rapid change in spectral
properties was a unique phenomenon, never before seen in any hot post-AGB star.

\section{Spectroscopic Monitoring of Lo 4}

At the time of the 1992 mass-loss episode, only a few spectra had ever been
taken of Lo~4. This suggested that the outbursts are likely not rare events, in
Lo~4 itself, and possibly in other PG~1159-type PNNi. 

In order to explore this possibility, I began a program of regular spectroscopic
monitoring of Lo~4 in early 2003, and continued it until early 2012.  I used the
1.5-m telescope operated by the SMARTS Consortium\footnote{SMARTS is the Small
\& Moderate Aperture Research Telescope System; {\tt
http://www.astro.yale.edu/smarts}}, located at Cerro Tololo Interamerican
Observatory. The 1.5-m telescope was equipped with a spectrograph at the RC
focus and a CCD camera. Data were obtained in queue-scheduling mode by Chilean
observers. I used the 26/I grating setup, giving a wavelength coverage of
3650--5400~\AA, with a resolution of 4.3~\AA\null. Exposure times were
$3\times900$~s, yielding a S/N per resolution element of $\sim$30--40 on good
nights.  The CCD images were bias-subtracted and flat-fielded, combined for
cosmic-ray removal, and then the spectra were extracted and
wavelength-calibrated, all using standard IRAF\footnote{IRAF is distributed by
the National Optical Astronomy Observatory, which is operated by the Association
of Universities for Research in Astronomy (AURA) under cooperative agreement
with the National Science Foundation.} routines.

Data were obtained at essentially randomly chosen dates. I have given progress
reports at two conferences (Bond 2008, 2010); the monitoring has now been
discontinued and I present the final results here. Between 2003 February and
2012 February, 81 usable spectra of Lo~4 were obtained. Table~1 lists the dates
of the observations. Four new mass-loss events were detected during this
interval. Table~2 summarizes the dates of these spectroscopic outbursts. 

Fig.~1 illustrates, as an example, the mass-loss episode that occurred on 2008
November~30.  The spectra have been normalized to a flat continuum, and a
3-point boxcar smoothing has been applied. The top spectrum, 19~days before the
event, is noisy but shows the normal state of Lo~4. On November~30 the
\ion{O}{6} doublet at 3811-3834~\AA\ had strengthened over its normal value, and
the \ion{O}{6} 5291~\AA\ emission line was much stronger than usual. Most
dramatic was the appearance in strong emission of the \ion{C}{4}+\ion{He}{2}
feature at 4659--4686~\AA, which is normally seen in absorption. This spectrum
is very similar to the outburst spectrum detected by Werner et al.\ in 1992. An
observation two days later, obtained on 2008 December~2 (not included in Fig.~1
because poor observing conditions made it very noisy), showed that the
\ion{C}{4}+\ion{He}{2} emission was already gone. A better spectrum from
December~4, shown at the bottom of Fig.~1, verifies the absence of
\ion{C}{4}+\ion{He}{2} emission, as well as the near-disappearance of the
\ion{O}{6} 5291~\AA\ emission line.

I created a high-S/N spectrum of Lo~4 in its low state by averaging the 58 best
spectra obtained between 2003 and 2012. This spectrum is shown at the top of
Fig.~2, with the most prominent features of \ion{He}{2}, \ion{C}{4}, and
\ion{O}{6} marked. The next four spectra in the figure show the appearance
during the four detected ``high'' states of greatly enhanced mass loss. The time
intervals between these episodes and the previous and following observations
showing normal low-state spectra were as follows: (1)~2006 Jan.~16: spectra 53
days earlier and 15 days later were normal; (2)~2006 Nov.~30: 56 days earlier,
17 days later; (3)~2008 Nov.~30: 19 days earlier, 2 days later; (4)~2011
Oct.~30: 164 days earlier, 21 days later. Thus, although the constraints are not
tight, the events are likely of relatively short duration, no more than a few to
several days, in agreement with the findings reported by Werner et al.\ (1992).

\section{Observations of the Central Star of NGC 246}

The nucleus of the PN NGC~246 is another PG~1159-type star, considerably
brighter than Lo~4 (Bond \& Ciardullo 1999 measured $V=11.8$). Its parameters
are similar within the errors to those of Lo~4; in the same paper in which they
discussed {\it FUSE\/} observations of Lo~4, Werner et al.\ (2007) reported
$T_{\rm eff}=150,000$~K, $\log g=5.7$ for NGC~246. A further indication of the
similarity of the two stars arises from the detection of GW~Vir-type non-radial
pulsations in NGC~246 (Ciardullo \& Bond 1996; Gonz{\'a}lez P{\'e}rez, Solheim,
\& Kamben 2006), with periods of $\sim$24--31~min.

Does NGC~246 also have occasional episodes of enhanced mass loss? To explore
this possibility, I obtained spectra with the SMARTS 1.5-m telescope for this
star on 167 occasions, between 2003 June and 2011 July. In addition to the 26/I
grating setup described above, the brightness of NGC~246 made it possible to use
two higher-resolution setups: 56/II (4017--4938~\AA, 2.2~\AA\ resolution) and
47/IIb (4070--4744~\AA, 1.6~\AA\ resolution).  Typical exposure times were
$3\times240$~s for 26/I and $3\times300$~s for 56/II and 47/IIb. In spite of the
similar spectra, stellar parameters, and pulsational properties of the two
stars, NGC~246 never showed mass-loss events like those of Lo~4 during the
observations between 2003 and 2011. I will provide the observation dates to any
interested readers.

I created a high-S/N spectrum for NGC~246 by averaging 21 excellent spectra.
Fig.~3 compares this spectrum with the average low-state spectrum of Lo~4, and
illustrates the similarity. However, the \ion{O}{6} emission features are
noticeably weaker in NGC~246.

\section{Constraints and Speculations}

I detected four mass-loss episodes during 81 observations of Lo~4. Following the
precepts used by Zwicky (1938) to determine the average rate of supernova
explosions per galaxy, I can estimate the typical time between outbursts of Lo~4
(on the assumption that the recurrence rate has been approximately constant over
the observing interval). Using Zwicky's terminology, the ``control value'' of a
single observation is the length of time during which an outburst would have
been detectable (i.e., the duration of a typical mass-loss episode). Based on
Werner et al.\ (1992) as well as my own results, the control value for Lo~4 is
about $\tau\simeq5$~days. Thus the total control value of the 81 observations is
$81\,\tau = 405(\tau/5)\,\rm days$ (neglecting the small number of observations
separated by less than $\tau$~days). Since 4 outbursts were detected, the mean
rate of mass-loss events in Lo~4 is found to be about one every
$101\,(\tau/5)\,\rm days$ (with a statistical uncertainty of about $\pm$50\%
since only 4 events were detected).  Moreover, we can estimate that the mean
fraction of the total elapsed time during which Lo~4 was in outburst is about
$\tau/[101\,(\tau/5)] \simeq 5\%$, independent of the outburst duration as long
as it is short compared to the typical spacing of the observations.

At present there seems to be no entirely satisfactory explanation for these
transient mass-loss episodes in Lo~4.  One constraint is that the temporary
stellar wind is {\it hydrogen-deficient\/} and {\it helium-rich}, and thus must
arise from photospheric material. This shows that the outbursts are unlikely to
be related to accretion events involving a companion star, debris disk,
infalling planet, or other exotic external cause. No 24~$\mu$m dust excess in
Lo~4 was found in a {\it Spitzer\/} survey by Chu et al.\ (2011). The short
durations of the episodes indicate that they are not driven on an evolutionary
timescale.

Lo~4 lies in the HR diagram for hot post-AGB remnants near the boundary between
emission-line [WCE] and predominantly absorption-line PG~1159 objects (e.g.,
Fig.~1 of Werner 2001). The general picture (cf.\ Koesterke \& Werner 1998;
Werner 2001; Koesterke 2001; Liebert 2008, Fig.~1) is that, as such stars evolve
to lower luminosity, the strong radiatively driven winds of [WCE] objects
gradually diminish, weakening the emission-line spectrum and transitioning the
star to a PG~1159 spectral type. This suggests that, for a PG~1159 star just
finishing this slow transition, it might take only a small perturbation to move
it back into the [WCE] stage. A possible speculation is that occasionally many
pulsation modes may be in phase, giving a temporarily large surface-temperature
amplitude, which might trigger the outbursts. If so, we might expect the
outbursts to occur periodically, at the beat period between the pulsation modes.
However, I could find no obvious period that would predict the outburst dates
listed in Table~1 without also predicting outbursts that were not seen to occur.

If we adopt the outburst mass-loss rate of $5\times10^{-8} \, M_\odot \, \rm
yr^{-1}$ given above, and if the star is in outburst $\sim$5\% of the time, the
integrated mass-loss rate from Lo~4 due to its outbursts is
$\sim$$2.5\times10^{-9}\,M_\odot\,\rm yr^{-1}$. There has not to my knowledge
been a published estimate of the mass-loss rate for Lo~4 in its normal low
state. However, a rough comparison is provided by the estimates for the similar
K~1-16 and NGC~246 of $\sim$$10^{-8}$ and $10^{-7} \, M_\odot \, \rm yr^{-1}$,
respectively (Koesterke \& Werner 1998).  Thus the outbursts of Lo~4 appear to
have relatively little additional effect on the evolution of the star.

We are left with a curious astrophysical puzzle: what is the origin of these
episodic events in Lo~4? It would be useful to monitor the spectra of other PNNi
lying near the [WCE]--PG~1159 transition. A stringent test of the hypothesis
that in-phase pulsations produce the Lo~4 outbursts would be to search for the
onset of an enhanced pulsation amplitude in the light curve and verify that a
spectroscopic event is then triggered; however, since we cannot predict these
outbursts in advance, such a program would be very costly in telescope time.

\acknowledgments

I thank the STScI Director's Discretionary Research Fund for supporting our
participation in the SMARTS consortium, Fred Walter for scheduling the queue
observations, and the expert Chilean service observers who obtained the spectra
during many long clear Tololo nights:
Claudio Aguilera,
Sergio Gonz\'alez,
Manuel Hern\'andez, 
Rodrigo Hern\'andez,
Alberto Miranda,
Rustum Nyquist,
Alberto Pasten,
and Jos\'e Vel\'asquez.
Klaus Werner provided encouragement and useful discussion and comments. I thank
Robin Ciardullo for pointing out Zwicky's classical paper.

{\it Facilities:} \facility{SMARTS 1.5-m telescope}


\begin{figure}
\begin{center}
\plotone{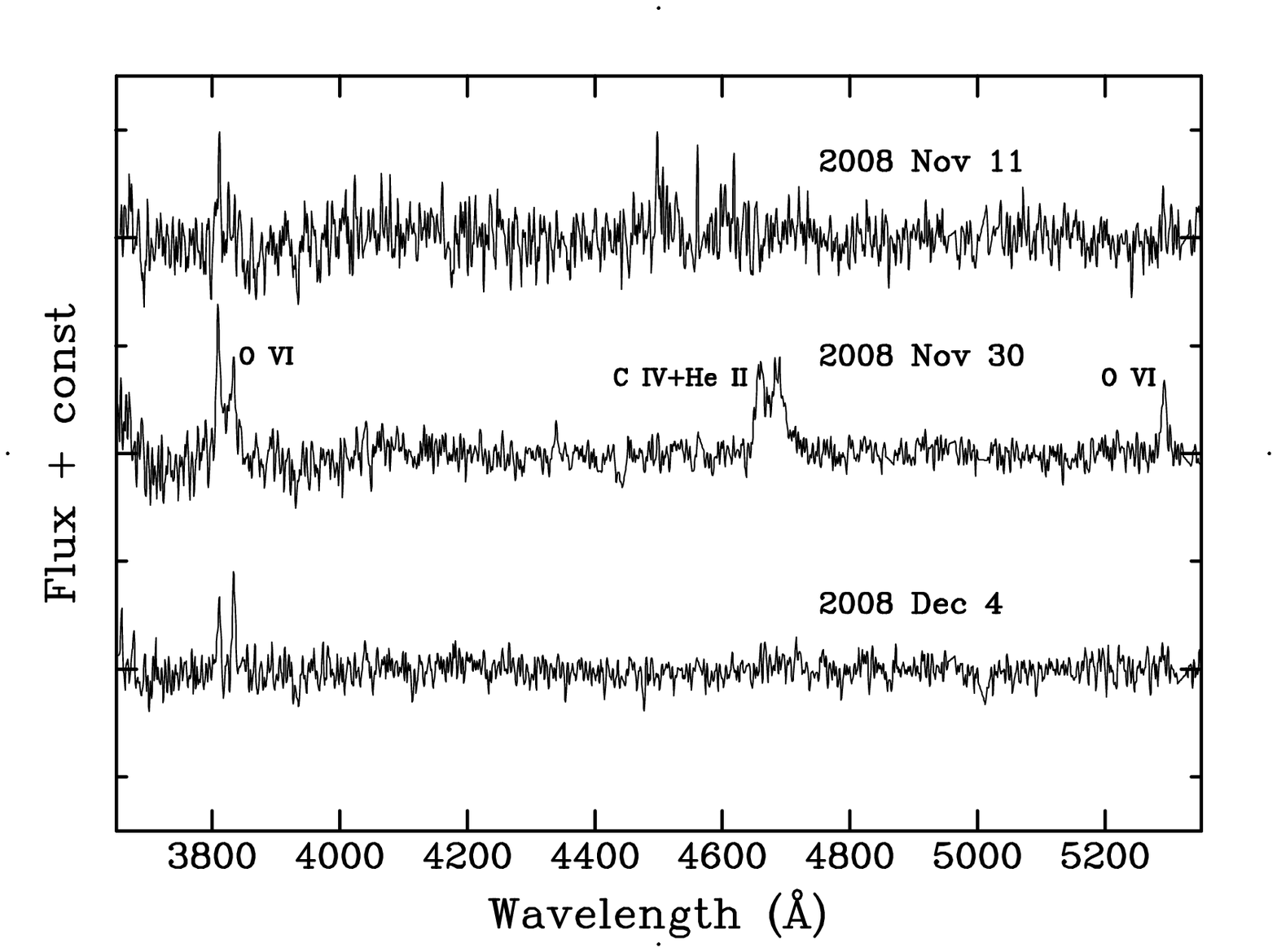}
\figcaption{
{\it Top:} SMARTS 1.5-m spectrum of Lo~4 on 2008 Nov~11; although noisy, it
shows the star in its normal quiescent state.
{\it Middle:} spectrum on 2008 Nov 30, showing that a mass-loss episode is
underway. The \ion{C}{4}+\ion{He}{2} feature is now strongly in emission, and
the emission at the \ion{O}{6} lines is also stronger than in the normal state.
{\it Bottom:} spectrum on 2008 Dec~4. Lo~4 is back to its normal state. All
spectra have been normalized to a flat continuum, and tick marks on the $y$-axis
are spaced at 0.5 of the continuum level.
}
\end{center}
\end{figure}

\begin{figure}
\begin{center}
\includegraphics[width=5.5in]{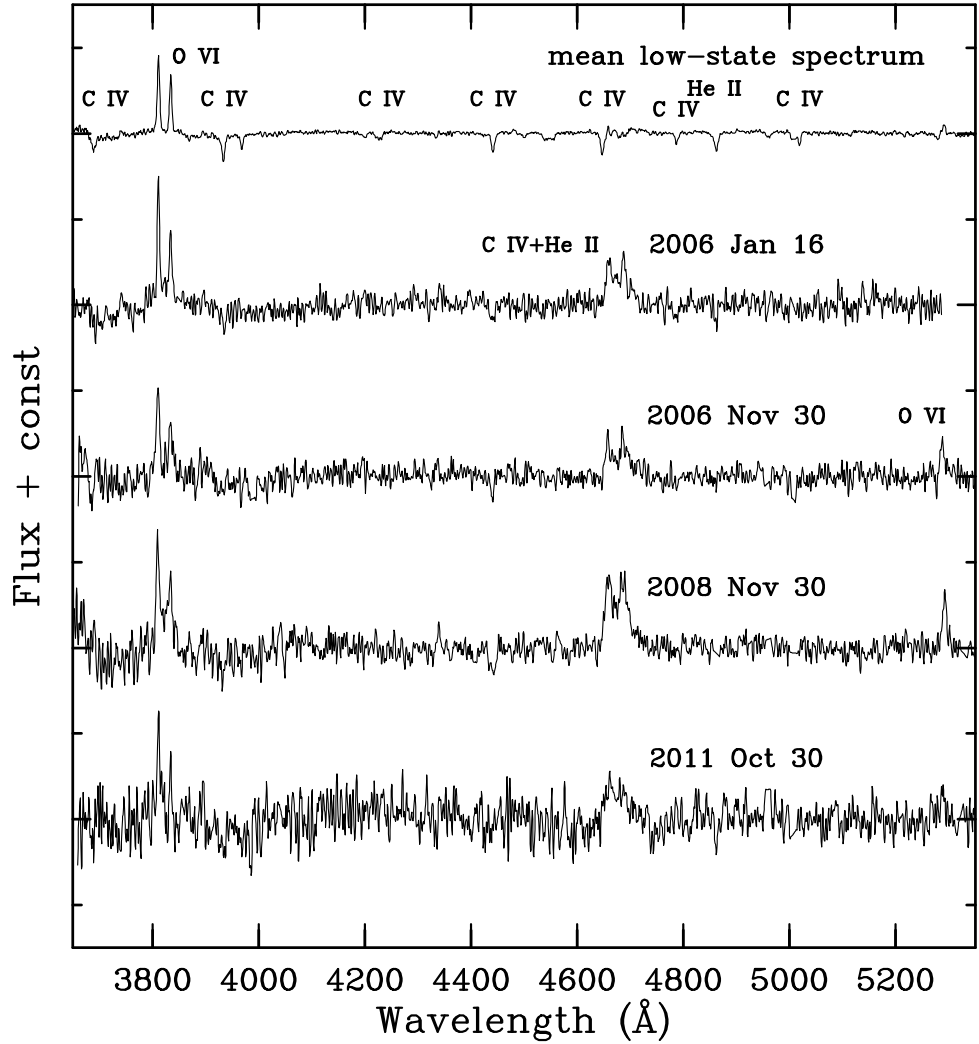}
\figcaption{
{\it Top:} high-S/N average spectrum of Lo~4 in its normal low state.
Prominent
absorption lines of \ion{He}{2} and \ion{C}{4} are labelled, along with the
emission doublet of \ion{O}{6} at 3811-3834~\AA\null.
{\it Bottom:} spectra of Lo~4 showing the four mass-loss events detected during
the 2003--2012 interval. 
Tick marks on the $y$-axis are spaced at 0.5 of the continuum
level.
}
\end{center}
\end{figure}

\begin{figure}
\begin{center}
\plotone{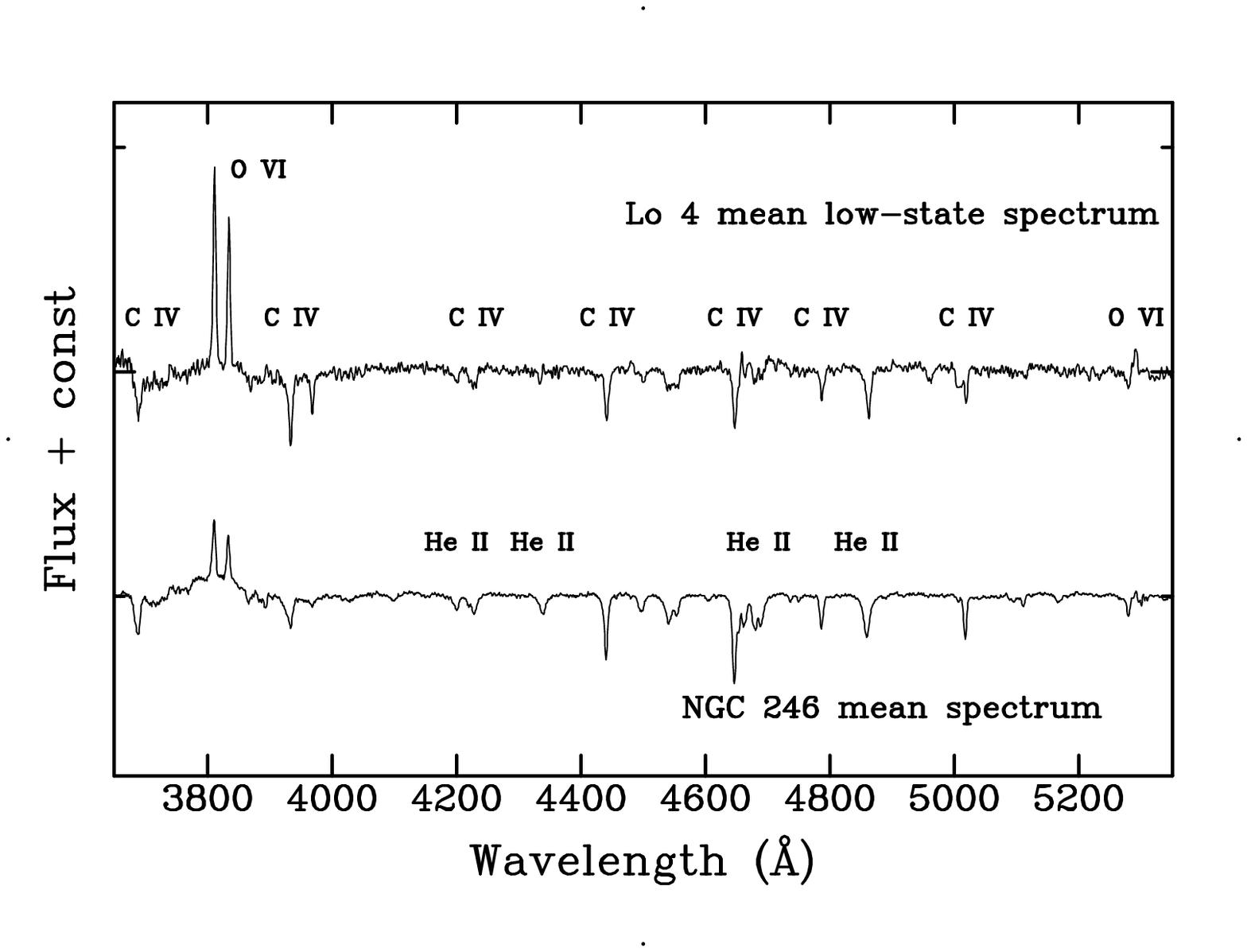}
\figcaption{
{\it Top:} high-S/N spectrum of Lo~4 in its normal state from Fig.~2.
{\it Bottom:} high-S/N spectrum of the central star of NGC~246.  Tick marks on the $y$-axis are spaced at 0.5
of the continuum level.
}
\end{center}
\end{figure}

\clearpage

\begin{deluxetable}{llll}
\tablewidth{0 pt}
\tablecaption{Dates of Spectroscopic Observations of Lo 4}
\tabletypesize{\footnotesize}
\tablehead{
\colhead{Civil Date} &
\colhead{Civil Date} &
\colhead{Civil Date} &
\colhead{Civil Date} 
}
\startdata\small							       
2003-02-19 & 2006-05-08 & 2008-05-15 & 2010-10-06 \\
2003-03-12 & 2006-06-03 & 2008-06-21 & 2010-11-10 \\
2003-04-04 & 2006-10-04 & 2008-07-03 & 2010-11-26 \\
2003-04-19 & 2006-11-29 & 2008-07-13 & 2010-12-29 \\
2003-05-07 & 2006-12-16 & 2008-10-23 & 2011-01-10 \\
2003-06-11 & 2006-12-18 & 2008-11-10 & 2011-01-27 \\
2003-06-25 & 2006-12-20 & 2008-11-29 & 2011-02-12 \\
2004-01-11 & 2006-12-22 & 2008-12-01 & 2011-03-02 \\
2004-01-31 & 2007-01-17 & 2008-12-03 & 2011-03-16 \\
2004-04-16 & 2007-02-08 & 2008-12-07 & 2011-04-16 \\
2004-11-05 & 2007-02-11 & 2008-12-28 & 2011-05-18 \\
2005-01-02 & 2007-04-11 & 2009-01-13 & 2011-10-29 \\
2005-01-15 & 2007-06-27 & 2009-01-28 & 2011-11-19 \\
2005-02-08 & 2007-07-18 & 2009-02-14 & 2011-11-29 \\
2005-05-06 & 2007-12-16 & 2009-02-26 & 2011-12-27 \\
2005-06-03 & 2007-12-30 & 2009-03-19 & 2012-01-05 \\
2005-11-23 & 2008-02-07 & 2009-04-14 & 2012-01-23 \\
2006-01-15 & 2008-02-25 & 2009-05-31 & 2012-02-14 \\
2006-01-30 & 2008-03-08 & 2009-12-16 & $\dots$ 	\\
2006-03-07 & 2008-03-29 & 2010-01-09 & $\dots$ 	\\
2006-04-02 & 2008-05-03 & 2010-05-23 & $\dots$  \\
\enddata
\tablecomments{Dates are for beginning of night in format YYYY-MM-DD; UT
dates are one day later.}
\end{deluxetable}

\begin{deluxetable}{lc}
\tablewidth{0 pt}
\tablecaption{Mass-Loss Events for Lo 4}
\tablehead{
\colhead{UT Date} &
\colhead{HJD$-$2400000} 
}
\startdata
1992 Jan 28\tablenotemark{a} & 48649.70 \\
2006 Jan 16 & 53751.66  \\
2006 Nov 30 & 54069.79  \\
2008 Nov 30 & 54800.78  \\
2011 Oct 30 & 55864.84  \\
\enddata
\tablenotetext{a}{Werner et al.\ 1992}
\end{deluxetable}

\end{document}